\documentclass[journal=ancac3,manuscript=article]{achemso}

\usepackage[version=3]{mhchem} 
\usepackage{xr}
\usepackage{svg}

\author{Esli Diepenbroek}
\author{Leon A. Smook}
\author{Sissi de Beer}
\email{s.j.a.debeer@utwente.nl}
\affiliation[University of Twente]
{Department of Molecules and Materials, MESA+ Institute, University of Twente, P.O. Box 217, 7500 AE, Enschede, The Netherlands}

\title{The role of polyelectrolyte brushes in tunable synaptic devices}

\abbreviations{}

\begin{document}

\begin{tocentry}
\includegraphics[width=8.25cm]{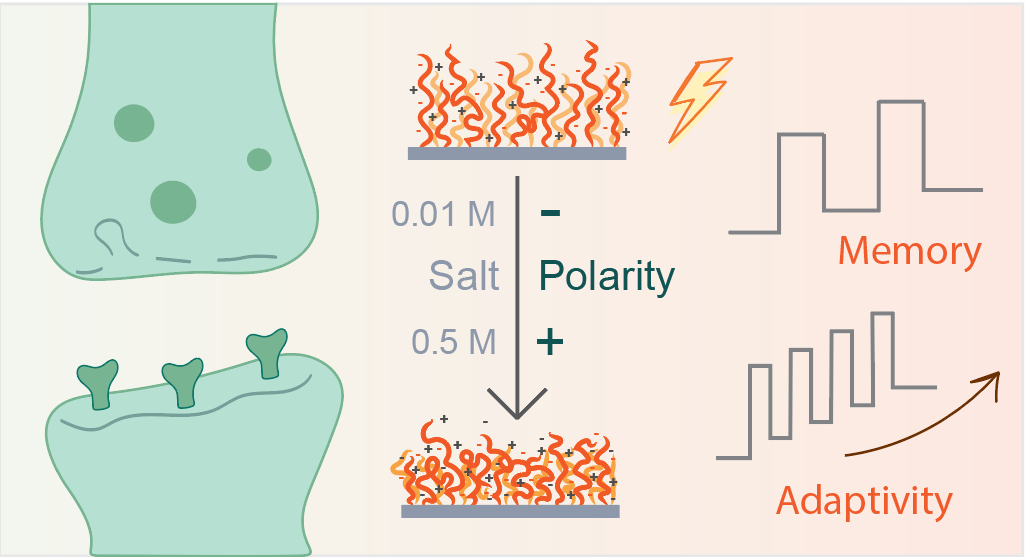}
\end{tocentry}

\begin{abstract}
With the ever-increasing digitization of society, the development of materials with low-power memory storage -- similar to synapses -- is becoming more relevant. The field of iontronic artificial synapses has gained traction, in particular with polymers as the memory-active material which allows for additional bio-compatibility, flexibility and tunability. Polyelectrolyte brushes are an example of stimulus-responsive materials that can be used in iontronic devices. However, the complexity of current neuromorphic devices does not allow us to isolate and understand the role of polyelectrolyte brushes in their synaptic response. In this paper, we show that polyelectrolyte brushes are capable of synaptic behavior in the most simple of electrochemical cell designs. Furthermore, by combining theory and experimental work, we shed light on the role of brushes in this synaptic behavior and their dynamic stimuli-responsiveness to polarity changes for different salt concentrations. The obtained trends and interpretations of the nonlinear potential-current response, paired‑pulse experiments, and accumulative learning lay the foundation for designing and developing polymer brush‑based neuromorphic devices. 
\end{abstract}


\section{Introduction}
Our brain is capable of high-density, low-power memory storage owing to the dynamic potentiation and depression of synapse potentials~\cite{Cai.Networks.2014,Seok.AdvElectrMater.2024,Huang.SciAdv.2024}. With the ever-increasing digitization of society and proliferation of artificial intelligence, the development of materials with similar brain-like functions is becoming more relevant~\cite{Wang.NatMater.2022,Zhang.ACSNano.2025,Seok.AdvElectrMater.2024,Ro.ACSNano.2025}. \textit{Artificial synapses} drive research progress in the field of neuromorphic devices, which find their use in neuromorphic computing~\cite{ Zolfagharinejad.EurPhysJB.2024,Zhang.ACSNano.2025,Robin.Science.2023,Hou.PhysChemLett.2023}, bionic devices~\cite{Xia.Nanoscale.2024}, power devices\cite{Kim.AdvSci.2023} and bio-electronic interfaces.\cite{Wu.AdvFuncMater.2025} Artificial synapses can emulate the dynamic behavior of biological synapses by exhibiting \textit{plasticity} -- the ability to remember a previous trigger -- as well as memristive behavior.~\cite{Chuo.Handbook.2019,Lin.AdvMateri.2013,Li.ACSApplBioMat.2020}. So far, most neuromorphic devices are based on conductivity switching by (opto)electronic signals~\cite{Zhang.ACSNano.2025,Li.ACSApplBioMat.2020,Qiu.ACSNano.2025,Luo.ACSNano.2019,Wang.NatMater.2022}. In such devices, semiconductors\cite{Luo.ACSNano.2019,Lee.ACSNano.2024} and polymer thin films\cite{Park.ACSNano.2017,Qiu.ACSNano.2025,Liu.Science.2025,Burgt.NatMater.2017} or organic molecular junctions~\cite{Zhang.NatComm.2024,Wang.NatMater.2022} are used to transport electronic information carriers. Alternatively, the biological process of neurotransmission can be mimicked by ion-current rectification,~\cite{Kamsma.Faraday.2026,Hu.SciAdv.2025,Robin.Science.2023,Kim.AdvSci.2023,Xu.ACSNano.2024,Huang.SciAdv.2024,Hou.PhysChemLett.2023,Li.ACSApplBioMat.2020,Li.ACSNano.2023} using ions as the information carrier. Unlike electrons, ions appear in different sizes and valencies, and facilitate different types of interactions, which makes their migration slower and their information density higher.\cite{Li.ACSApplBioMat.2020} As such, the field of iontronic synapses has gained traction, in particular with polymers as the memory-active material that allow for additional benefits of bio-compatibility, flexibility, low-energy consumption and tunability\cite{Li.ACSApplBioMat.2020,Ro.ACSNano.2025,Xia.Nanoscale.2024,Yu.Small.2024,Burgt.NatMater.2017,Wu.AdvFuncMater.2025}.

Polymer-based, iontronic artificial synapses exhibit plasticity by either 1) using (redox) ionic--electronic coupling\cite{Burgt.NatMater.2017,Liu.AdvMater.2026,Mu.AdvFuncMater.2023}, 2) ion doping\cite{Yu.Small.2024,Hu.AdvFuncMater.2021,Li.ACSApplBioMat.2020} or by 3) ion transport through their polymer matrix\cite{Lee.AdvMater.2024,Wu.AdvFuncMater.2022,Wu.AdvFuncMater.2025}. Most synaptic devices with redox-active or ion-doped polymers require dedicated organic synthetic procedures\cite{Liu.AdvMater.2026,Mu.AdvFuncMater.2023} or additional polymer modification.\cite{Hu.AdvFuncMater.2021,Yu.Small.2024} Simple, charged polymers are used in iontronic synapses that are driven by ion transport. However, they combine these simple polymers with complex architectures to drive ion transport, including hierarchical channels~\cite{Hu.SciAdv.2025,Lee.AdvMater.2024,Wu.AdvFuncMater.2022,Allegretto.ACSNano.2024} or conical channels,~\cite{Kamsma.PRL.2023,Xiong.Science.2023} or ion-selective barriers~\cite{Kim.AdvSci.2023,Wu.AdvFuncMater.2022}. Sometimes these complex architectures include stimuli-responsive polymers that can adjust ion transport when an electrical signal is applied.~\cite{Zhang.ACSNano.2025}.

Polyelectrolyte brushes -- surface functionalizations with densely end-grafted polyelectrolyte chains~\cite{Smook.AdvMater.2025,Wang.AdvMater.2023} -- are an example of stimulus-responsive materials that can tune ion transport. Due to their high density of chains and covalent grafting, brushes impart surfaces with unique abilities, such as low-friction and antifouling properties.\cite{Feng.NanoLetters.2026,Blau.AdvHealthMater.2024,Smook.AdvMater.2025,Visova.AdvMaterInt.2022} As such, brushes have been used as a functional material in iontronic devices. They show memristive characteristics,\cite{Wolski.JCollIntSc.2023,Wolski.Small.2024} increase the capacitance of flat electrode surfaces~\cite{Qing.ACSNano.2023,Blau.AdvHealthMater.2024} and improve the performance of neural interfaces.\cite{Cho.ActBiomater.2024} When exposed to electrical potential differences, the structure of a polyelectrolyte brush changes~\cite{Weir.Langmuir.2011,Senechal.Macromolecules.2022,Senechal.SoftMatter.2020,Borisova.Langmuir.2015,delCastillo.Angewandte.2022} and molecular simulations reveal that brushes restructure through the collapsing or stretching of a fraction of the chains~\cite{Ouyang.Nanotechnology.2009,Ho.Langmuir.2013,Wang.ChemicalPhysics.2020,Smook.Langmuir.2024,Smook.Macromolecules.2025,Pial.Macromol.2022}. In addition, the restructuring is qualitatively affected by the presence of salt~\cite{Okrugin.Polymers.2020, Smook.JPolymSci.2025}. Typical response times of these systems are milliseconds to seconds~\cite{Drummond.PRL.2012,Borisova.Langmuir.2015,Weir.Langmuir.2011, Senechal.Macromolecules.2022}, and time-resolved sequential measurements indicate that brush dynamics show hysteresis, tunability and temporary irreversibility~\cite{Senechal.Macromolecules.2022}. We hypothesize that these brush dynamics are indicative of an intrinsic, volatile memory that could enable synaptic behavior. 

However, existing studies focusing on polymer-brush based artificial synapses do not elaborate on the role of brushes in synaptic behavior. Xiong \textit{et al}.\cite{Xiong.Science.2023} show synaptic behavior of polycation brushes in a conically shaped fluidic system. This study is the only work -- as far as we know -- that features synaptic plasticity experiments with polymer brushes, which shows that synaptic behavior can be influenced by the solvation state of the brush. However, similar conically shaped systems showed synaptic plasticity in the absence of polymer brushes\cite{Kamsma.PRL.2023}. Polymer brushes thus enable tuning of an artificial synapse, but do they display synaptic behavior regardless of their substrate geometry?

\newpage
In this paper, we show that polyelectrolyte brushes are capable of synaptic behavior, and we highlight the role of brushes in this synaptic behavior. The combination of experiments, mean-field approximations and molecular dynamics allow us to gain a mechanistic picture of the interplay between brushes and ions in response to changes in electrical potential. We also show that the brush response and the ion dynamics within the brush is asymmetric with respect to the polarity of the electric field and with the electrolyte salt concentration. The obtained results help open up the field of polymer brush-based neuromorphic devices, and lay the foundation for designing of future fluidic, biocompatible electronic interfaces.  

\newpage
\section{Results and discussion}
\begin{figure}
    \centering
    \includegraphics[width=.95\linewidth]{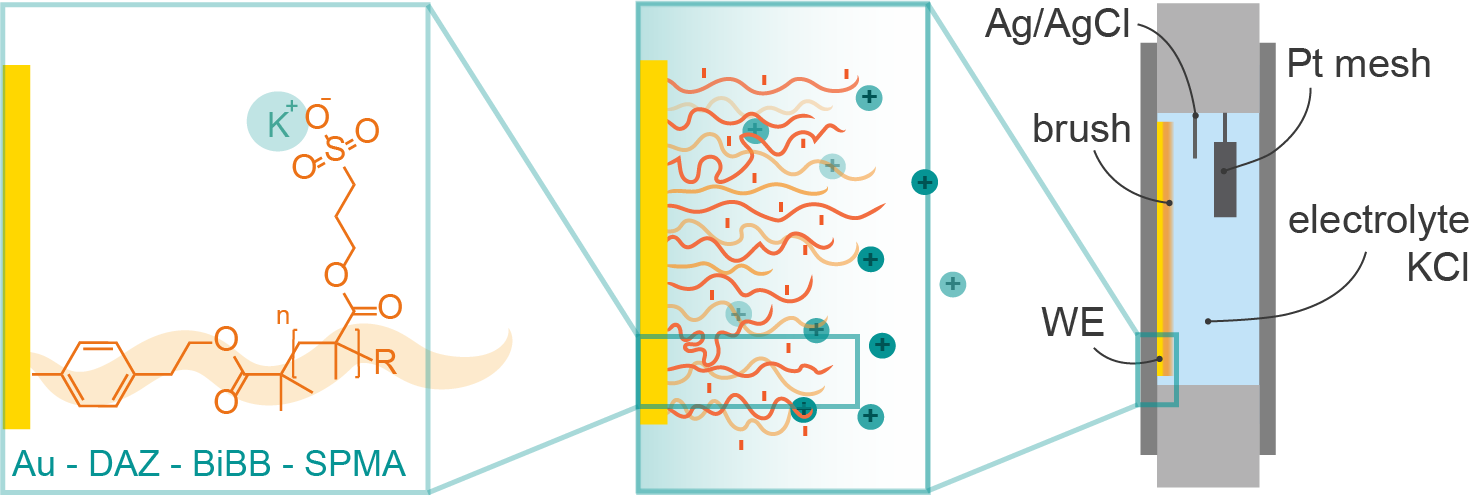}
    \caption{\textbf{Polyelectrolyte brushes on gold.} A schematic representation of our material; from the molecular structure of our PSPMA brushes (left) to a microscopic side-view of these brushes (middle), which is mounted in an electrochemical cell as a coated gold WE (right).}
    \label{fig:Materials}
\end{figure}

To enable synaptic behavior, we considered brushes made from a strong polyelectrolyte material. Strong polyelectrolyte brushes are promising candidates for showing short term plasticity, owing to their selective ion-permeability and electro-responsiveness in a broad pH-range.\cite{Smook.AdvMater.2025,Zhang.Langmuir.2017} For this reason, we synthesized polyanionic 3-sulfopropylmethacrylate (PSPMA) brushes on gold electrodes (\textbf{Figure \ref{fig:Materials}}) using atom-transfer radical polymerization (ATRP). The polymer strands are covalently bound to the electrode by a phenyl-based carbon to gold anchoring strategy ([C\textsubscript{ph}--Au]) developed in our lab.\cite{Postma.AdvMaterInt.2026} A step-by-step procedure of the PSPMA brush synthesis is included in the \textit{Supporting Information}. With this polymerization method, we reliably obtained brushes with a dry height of 14.0 $\pm$ 1.5 nm after 45 minutes and a grafting density of 0.19--0.22 chains/nm$^2$  (\textit{Supporting Information}, Section 1) This brush thickness far exceeds the thickness of the electric double layer $\lambda_D$ at the gold-brush interface (EDL; $\lambda$ = 0.4--3.0 nm); the brush will therefore play a dominant role in the observed potential-current coupling. The strong, covalent attachment to gold, combined with the chosen brush height,\cite{Postma.Macromol.2026} yields electrochemically stable PSPMA brushes. The polymer-brush functionalized gold substrates were mounted into an electrochemical cell with a three-electrode configuration (Figure \ref{fig:Materials}), using a Pt mesh as a counter electrode (CE) and Ag/AgCl as the reference electrode (RE). The PSPMA brushes on gold served as the working electrode (WE), which were solvated in a dilute (0.01 M) or concentrated (0.5 M) KCl electrolyte. 

The choice of salt concentration relates to the behavior of polyelectrolyte brushes in presence of salt (Figure \ref{fig:Swelling}). At 0.01 M KCl, the brush will reside in the \textit{osmotic regime}, characterized by the osmotic pressure driven swelling of the brush. At 0.5 M KCl, the presence of ions will largely screen the charges on the polymer chains, marking the quasi-neutral \textit{salted regime.}\cite{Smook.JPolymSci.2025,Pincus.Macromolecules.1991} As the chains effectively appear neutral in this salted regime, the chains will experiences less electrostatic repulsion, which is an effect that usually enhances brush swelling. As a result, the PSPMA brushes will be collapsed more when solvated in high-salt conditions. \newpage A visual representation of both brush regimes is included in \textbf{Figure \ref{fig:NonLinear}a}; the osmotic regime is characterized by a stretched out brush, while salted brushes are (partially) collapsed. By probing these two brush behaviors, we effectively change the resistive-capacitive properties of the brush layer.\cite{Smook.Macromol.2026} To understand what brush parameters are affected, we use the electrochemical approximation of a polymer brush as in \textbf{Figure \ref{fig:NonLinear}b}.\cite{Smook.Macromol.2026,Anthi.BiomaterSc.2021} Here, the PSPMA brush system is represented as an equivalent circuit that consists of a bulk solution resistance (\textit{R}\textsubscript{s}), brush polarization resistance (\textit{R}\textsubscript{b}), brush capacitance (\textit{C}\textsubscript{b}) and double layer-related constant phase element (\textit{Q}\textsubscript{DL}). When the salt concentration is varied, we vary the extent of brush swelling/collapsing and thus the brush resistance \textit{R}\textsubscript{b}, as well as the electric double layer and thus non-ideal capacitance \textit{Q}\textsubscript{DL}.\cite{Smook.Macromol.2026} We derived that the brush capacitance \textit{C}\textsubscript{b} is dominated by the charges on the polymer backbone itself,\cite{Smook.Macromol.2026} which implies that \textit{C}\textsubscript{b} will likely remain stable across different salt regimes. Considering these parameter influences, we hypothesized that salt concentration has a large influence on the transient current response and thus synaptic behavior of polyelectrolyte brushes.

\newpage
\subsection{Potential-current characteristics}

\begin{figure}
    \centering
    \includegraphics[width=\linewidth]{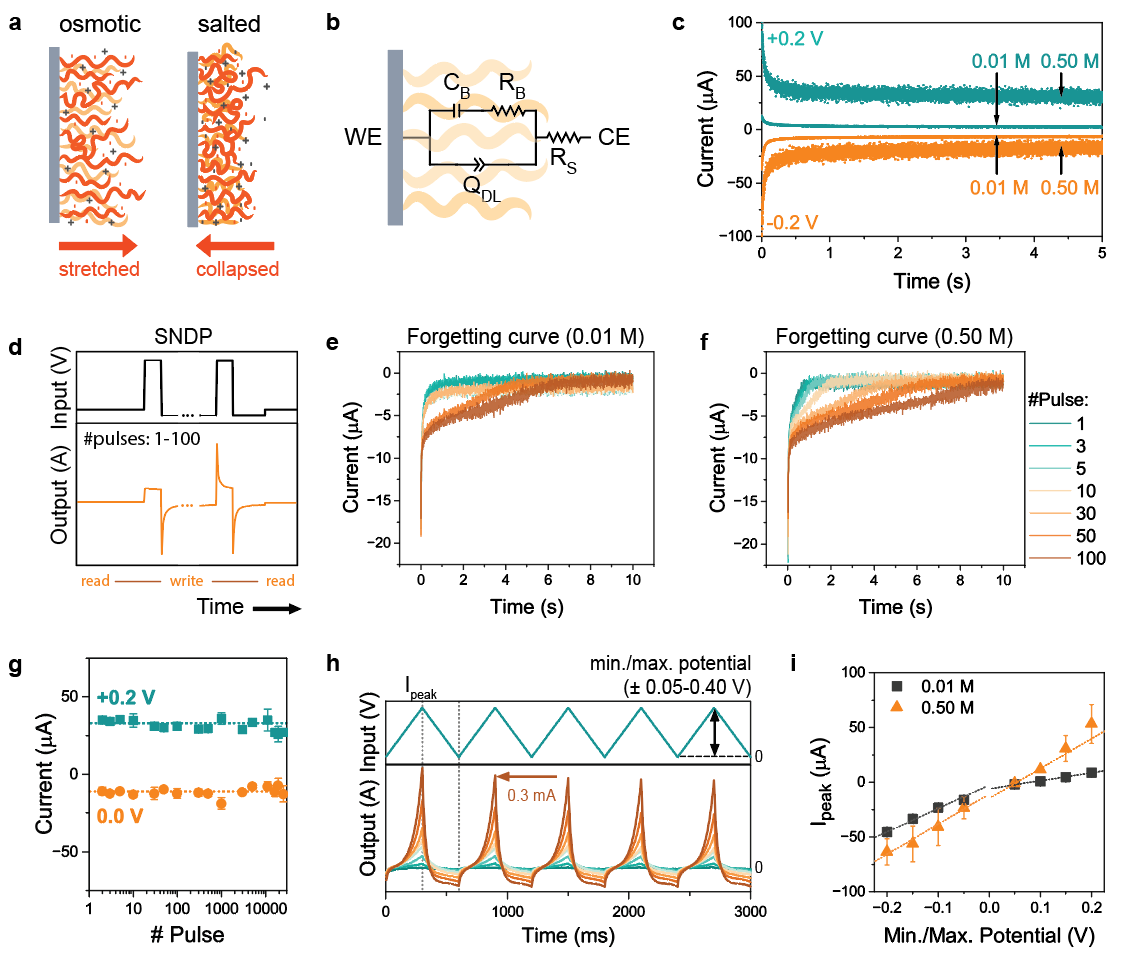}
    \caption{\textbf{Nonlinear responses of PSPMA brushes.}  a) Visual representation of an osmotic (left) and salted (right) brush. b) Equivalent circuit of a PSPMA brush in a KCl electrolyte solution. c) Chronoamperometry in osmotic and salted conditions, with polarities $\pm$ 0.2 V. d) Potential (input) and current (output) trace of a spike-number dependent plasticity (SNDP) experiment. Forgetting curves of an osmotic (e) and salted (f) brush. Pulse numbers are varied between 1-100 pulses. g) Stability of the PSPMA brushes across $>$25.000 pulses. h) Potential (input) and current (output) traces to assess the non-linear ion rectification behavior and potential-current coupling. i) Peak current as a function of the potential amplitude, as derived from (h).}
    \label{fig:NonLinear}
\end{figure}

Synapses are characterized by a specific set of features that allow them to \textit{learn} and \textit{forget}. Most characteristic is the appearance of synaptic plasticity, which refers to memory of a previous trigger at short retention times (milliseconds--minutes) or at long retention times (hours--days).\cite{Li.ACSApplBioMat.2020} Before we discuss the plasticity of our polyelectrolyte brushes in different salt environments, we first present various coupling characteristics between the applied potential (input) and measured current (output) in \textbf{Figure \ref{fig:NonLinear}c-i}. These characteristics reveal relevant transient processes that may play a role in the observed plasticity effects. 

\textbf{Figure \ref{fig:NonLinear}c} shows the step-response of our brush-functionalized interface when switching the potential from \textit{U = }0.0 V to $\pm$ 0.2 V in the osmotic and salted regime. The trends follow an experimental fit with two time constants (\textit{U}\textsubscript{0} + \textit{A}\textsubscript{1} $e^{-t/\tau_1}$ + \textit{A}\textsubscript{2} $e^{-t/\tau_2}$). We attribute the appearance of transient process 1, with time constant $\tau_1$, to double layer formation/relaxation connected to the non-ideal capacitor \textit{Q}\textsubscript{DL} in \textbf{Figure \ref{fig:NonLinear}b}. The assignment of the shorter time-scale $\tau_1$ to EDL-dynamics makes sense intuitively, as ions tend to react and mobilize faster than long, sterically hindered, polymer chains. Brushes solvated in 0.01 M KCl display a wider variety in EDL relaxation times ($\tau_1 =$ 33-53 ms) than when solvated in 0.5 M KCl (\textbf{$\tau_1 =$ }34-35 ms). Also, we observed a trend of $\tau_{1,+0.2 V} > \tau_{1,-0.2 V}$ in the osmotic regime, while in 0.5 M KCl, this difference becomes negligible. We attribute the time constant $\tau_2$ to brush relaxation, aligned with the brush capacitance \textit{C}\textsubscript{b} ($\tau_2$ = 0.63--0.69 s). We based the assignment of constants $\tau_1$ and $\tau_2$ on the calculation of $R_b [\Omega] \times C_b [s/\Omega]$, which yields values in the same order of magnitude as $\tau_2$ (0.19--0.49 s). In addition, reference measurements on non-coated gold WE yield one time constant, which is in the same order of magnitude as $\tau_1$ (37--58 ms). An overview of the time constants can be found in the \textit{Supporting Information}, Table \ref{tab:tauref}-\ref{tab:tau}.

The described trends in relaxation suggest that the brush material should facilitate memory-like behavior at timescales smaller than $\tau_1$, and more interestingly, that this behavior is salt and polarity-dependent. To illustrate the effect of salt in more detail, we tested the spike number dependent plasticity (SNDP: \textbf{Figure \ref{fig:NonLinear}d-f}) at positive potential in the osmotic and salted regime. During these SNDP experiments, the system is subjected to +0.2 V pulse trains of varying pulse numbers (here: \textit{\# pulses:} 1-100). By \textit{reading} the current after this pulse train with a reading voltage +20 mV, we monitor the forgetting behavior and thus the retention time of ions in the polymer brush. By varying the pulse numbers, as we do in such SNDP experiments, we can verify whether charge separation is enhanced by extending the sequence of pulses, which is characteristic of synaptic plasticity. We expand on such SNDP experiments in the next section, Figure \ref{fig:SRDP}b. We used a fixed pulse train frequency of 100 Hz, to ensure the inter-pulse time $\Delta t$ is shorter than the relevant time constant $\tau_{1,+0.2 V}$. The current traces in \textbf{Figures \ref{fig:NonLinear}e-f}, which show the reading phase (at +20 mV) after the pulse train, reveal two important trends. In the first place, we confirm that the polymer brush-coated WE is indeed able to store information and that the retention of the stored information scales with the pulse number. This is evident by the relaxation of the forgetting curves, which takes longer with increasing pulse number in both salt regimes. The second observation is that this delayed relaxation is more extreme at 0.5 M. The results in 0.5 M even suggest that at pulse numbers $>$ 100, the retention time extends beyond 10 seconds, and thus long term memory (LTM) behavior may be achieved. As the values of $\tau_1$ appear similar across both salt regimes, seeing a stronger effect of pulse number in the salted regime may seem counter-intuitive. In particular, it suggests that more factors contribute to the memory-like behavior that is not directly captured by the ion transport in brushes, related to $\tau_1$; we will explore this in the later sections. We also confirmed stable resistive switching of our PSPMA brushes by repeatedly exposing the coated WE to +0.2 V pulses (t\textsubscript{p}=10~ms, $>$ 25.000 pulses). \textbf{Figure \ref{fig:NonLinear}g} presents the resistive switching between \textit{U = } 0.0 V (off) and \textit{U =} + 0.2 V (on) of a PSPMA brush solvated in 0.01 M KCl, confirming stability throughout extensive use. The 0.5 M KCl equivalent (Figure \ref{fig:Stability}) shows deviating behavior at exposures exceeding $>$ 5000 pulses; this could be indicative of brush degrafting\cite{Ding.JMaterChemB.2022,Postma.Macromol.2026} or long term memory. To elucidate on the role of degrafting, we tested the brush thicknesses before/after the pulse trains, which shows that the brushes largely retain their thickness in both salt regimes (98.3 \% and 79.1\% for 0.01 M and 0.5 M, respectively). The combination of thickness variations and change in resistive switching suggests that in high salt environment and repeated exposures, we likely observe a dual effect of 1) degrading PSPMA chains and 2) transitioning from short- to long term plasticity. 

Next, we investigated the potential-current coupling of our PSPMA brushes in more detail, with the twofold purpose to 1) monitor the dynamic resistive-capacitive behavior of our brushes and 2) determine a suitable potential window for subsequent synaptic tests. To assess this coupling between potential input and current output, we subjected our brushes to a constant-phase triangular potential wave from 0 V to varying minimum/maximum potentials of $\pm$ 50--400 mV (\textbf{Figure \ref{fig:NonLinear}h}). We traced the current output over time and extracted the current at the amplitude \textit{I}\textsubscript{peak} to construct \textbf{Figure \ref{fig:NonLinear}i}. The current trace from \textbf{Figure \ref{fig:NonLinear}h} reveals a nonlinear potential-current behavior at higher amplitudes ($>$ 200 mV), characteristic of resistive-capacitive systems, as we have illustrated in \textbf{Figure \ref{fig:NonLinear}b} and characterized in Figure  \ref{fig:Synthesis}e and Table \ref{tab:ec_params}. The peak current shows a linear relation with min./max. potentials of $\pm$ 50--200 mV, indicating that faradaic contributions may start playing a significant role at higher amplitudes (\ref{fig:Nonlinear04}). The established linear relation once more fortifies the resistive-capacitive behavior that is illustrated in \textbf{Figure \ref{fig:NonLinear}b}.\cite{Anthi.BiomaterSc.2021,Smook.Macromol.2026} Interestingly, the slopes in \textbf{Figure \ref{fig:NonLinear}i} change with polarity for 0.01 M (0.21 $\pm$ 0.01 vs. 0.07 $\pm$ 0.00) and remain constant for 0.5 M (0.27 $\pm$ 0.03 vs. 0.27 $\pm$ 0.03). The change in slope at 0.01 M indicates a polarity-dependent conductivity, which may arise from a significant difference in migrating ion type, ion mobility or brush configuration. We set out to gain a mechanistic understanding of the brush plasticity and its stimuli-responsiveness in the next section.

\subsection{Non-volatile memory of polyelectrolyte brushes}

\begin{figure}[ht!]
    \centering
    \includegraphics[width=\linewidth]{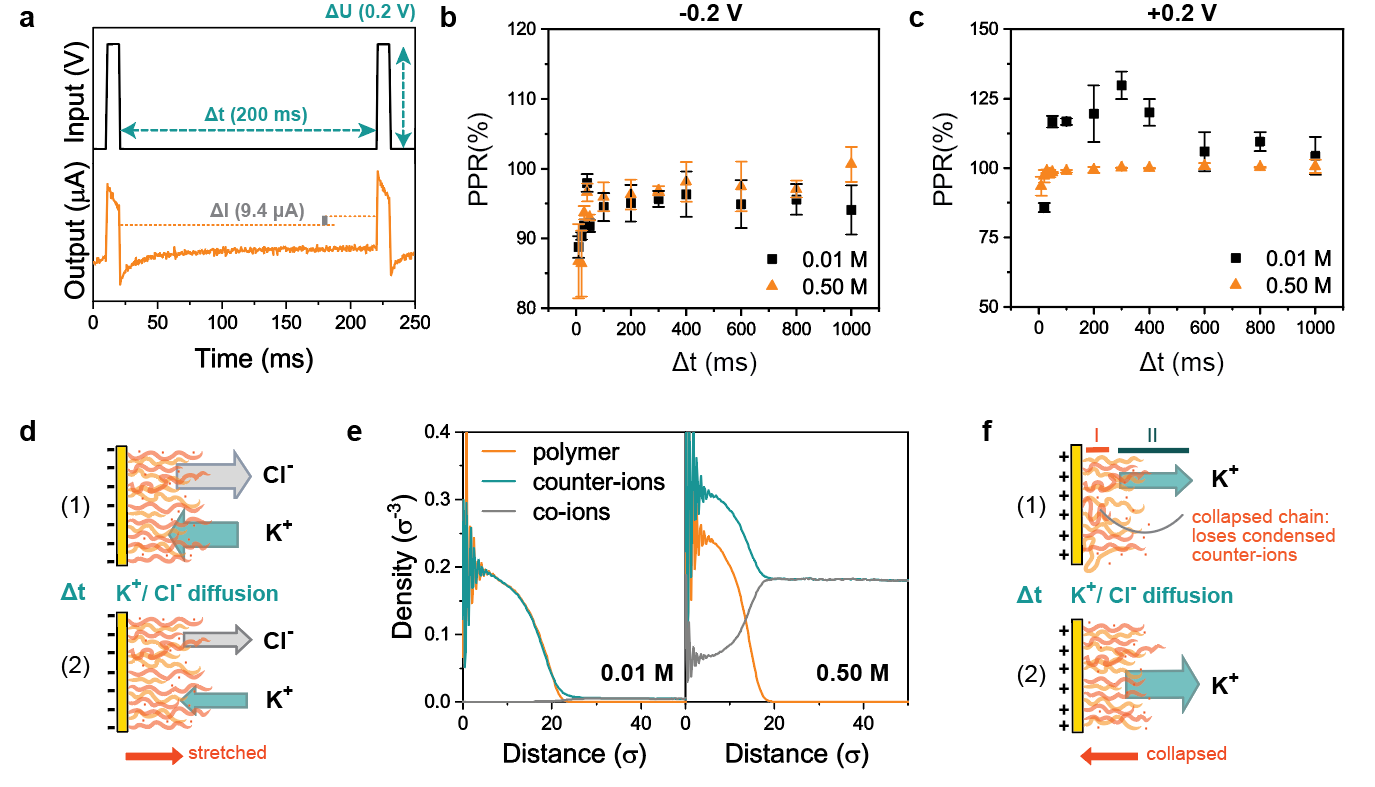}
    \caption{\textbf{Paired-Pulse Characteristics.} a) Time trace of a paired-pulse experiment with $\Delta$t = 200 ms. Paired-pulse experiments using pulses of \textit{U} = -0.2 V (b) or \textit{U} = +0.2 V (c) with PSPMA brushes in the osmotic (0.01 M) and salted (0.5 M) regime. d) Mechanistic picture of the trends in (b). e) Ion distributions along the z-direction using MD simulations of a representative charged brush (see \textit{Supporting Information}). f) Mechanistic picture of the trend in (c), specifically the osmotic regime. Here, region I indicates the area close to the WE that is rich in \ce{K+} counter-ions due to the collapse of PSPMA chains, while region II comprises the area that will be depleted of \ce{K+} counter-ions upon applying a positive bias. }
    \label{fig:PP}
\end{figure}

Short-term learning and forgetting based on a single, presynaptic signal can be tested through paired pulse (PP) experiments. We exposed our system to two short pulses (\textit{t}\textsubscript{p}=10~ms, \textit{U}=$\pm 0.2$~V) with an interpulse distance ($\Delta$\textit{t}) ranging from 10 ms to 1 s. By comparing the magnitude of the response current between the pump and probe pulse (\textbf{Figure \ref{fig:PP}a}), we can evaluate whether the brushes display paired-pulse facilitation (PPF; $>100\%$) or paired-pulse depression (PPD; $<100\%$). Here, the terminology of \textit{facilitation} and \textit{depression }refer to the local concentration of ions near the WE; an increase in local ion concentration facilitates an increase in current (PPF), while a depletion of the ions suppresses the current (PPD). \textbf{Figures \ref{fig:PP}b-c} show the results of the paired pulse experiments sorted by the polarity of -0.2 V (b) and +0.2 V (c). The magnitude of the response current was quantified using the paired pulse ratio (PPR), defined as the ratio of the average current measured between 8–10 ms of the 10 ms probe pulse (\textit{I}\textsubscript{2}) to that of the pump pulse (\textit{I}\textsubscript{1}): \textit{PPR (\%)} = $I_2/I_1 \times 100\%$. 

From the paired-pulse experiments with PSPMA brush-coated WEs, we derive that in both salt regimes and for -0.2 V pulses (Figure \ref{fig:PP}b) there is a PPD trend at short-to-medium $\Delta$t values ($<$ 600 ms). The retention times are similar to the work of Xiong \textit{et al}.\cite{Xiong.Science.2023}, who showed that brush-functionalized surfaces retain memory (\textit{PPR} $\neq$ 100\%) up to 400-800 ms depending on the polarity of the pulses. The power consumption $W = \int U \cdot I dt$ amounts to 1.27 $\mu$J for osmotic and 26.97 $\mu$J for salted PSPMA brushes; while higher than most artificial synapses,\cite{Li.ACSApplBioMat.2020} our large exposed surface (0.8 cm\textsuperscript{2}) cannot be compared to nanometer-sized synaptic devices. When comparing the paired-pulse trends of our PSPMA brushes and bare gold substrates, we see that the PPD trend for brushes is maintained longer (up to $\Delta$t-values of 600 ms) than for bare gold (up to $\Delta$t-values of 50 ms; Figure \ref{fig:GoldPP}). In \textbf{Figure \ref{fig:PP}b}, no significant difference is observed between the PPR-values of the osmotic and salted regime, which is in line with similar slopes in \textit{I}\textsubscript{peak} vs. \textit{Min./Max. Potential} that we present in \textbf{Figure \ref{fig:NonLinear}i}; if the resistive-capacitive behavior is similar, we can expect similar memory-dependent behavior. \textbf{Figure \ref{fig:PP}d} provides a mechanistic picture of the role of the polyelectrolyte brush during the paired-pulse experiments with \textit{U} = -0.2 V. During the pump pulse,  \ce{K+} ions will be concentrated at the WE, while similarly charged \ce{Cl-} ions will be repelled. The relative amount of mobilizable \ce{K+} and \ce{Cl-} ions will be different in the osmotic and salted regime; this will be explored in the next paragraph. Simultaneously, there might be a slight stretching effect of the PSPMA brush from the electrode surface due to electrostatic repulsion. This stretching, however, we do not expect to be significant based on our tested (mild) potentials.\cite{Smook.JPolymSci.2025,Weir.Langmuir.2011} In the subsequent inter-pulse period ($\Delta$t), ions will diffuse and re-establish the initial ion distribution. However, if $\Delta$t is much smaller than $\tau_2$ , that means that the system is not yet equilibrated, with the area close to the brush interface still rich in \ce{K+} and depleted in \ce{Cl-}. As a result, during the probe pulse, the recorded current -- which is created by ions migrating to/from the electrode -- is less than during the pump pulse. The direct relation to ion diffusion implies that the PPD should be most apparent at the shortest time intervals, as many have shown in their work on memristive coatings;\cite{Xiong.Science.2023} this also what we observe. 
 
A similar phenomenon is observed in the salted regime at +0.2 V (Figure \ref{fig:PP}c); a PPD trend at short $\Delta$t values ($<$ 600 ms) toward equilibration. While the electrostatic forces are opposite, the mechanistic picture of depletion and delayed replenishment is similar. On the other hand, we observe a curious trend at 0.01 M KCl and +0.2 V, which reports a PPD at $\Delta$t $<$ 50 ms followed by a long-term PPF region (120-130 PPR\%) that start to relax at $\Delta$t=600--1000 ms. The system is fully recovered beyond 1000 ms (Table \ref{tab:LongTermPP}). To understand the salt-dependent plasticity behavior, we conducted molecular dynamics (MD) simulations in osmotic and salted conditions, which provided us with ion density values as a function of the distance $\sigma$ from the substrate (\textbf{Figure \ref{fig:PP}e}; methodology provided in the \textit{Supporting Information}). Here, we observe differences in the dominating ion species at small distances from the gold-brush interface: in the salted regime, both counter- (\ce{K+}) and co-ions (\ce{Cl-}) are present, while the co-ion concentration in the osmotic regime can be neglected. At positive potentials, we should also consider the role of our negatively charged PSPMA brushes. At low salt concentrations, the charge centers on the polymer chain are less screened, allowing some of these chains to collapse onto the electrode surface.\cite{Yamamoto.EPL.2011} At high salt concentrations, this phenomenon is prevented by the screening of the charge centers by the ions present in the electrolyte.\cite{Smook.JPolymSci.2025} Combining the ion distributions from \textbf{Figure \ref{fig:PP}e} and the effect of brush collapse, we can mechanistically explain the paired-pulse trend in the osmotic regime, with positive potentials, according to \textbf{Figure \ref{fig:PP}f}. During the pump pulse, the few present \ce{Cl-} ions will be concentrated at the WE, while \ce{K+} close to the brush interface (see figure: region II) will be repelled. In addition, there is an effect of the negatively charged PSPMA chains; some chains will collapse at the positive charged WE surface (in figure: region I) and increase the local polymer density, likely releasing any condensed counter-ions that were present on their polymer backbone.\cite{Yi.Macromol.2026} At short $\Delta$t times, when $\Delta$t$<\tau$\textsubscript{1}, the ions do not have enough time to diffuse back to region II: a PPD trend is observed. At intermediate $\Delta$t times, when $\tau$\textsubscript{1}$<$$\Delta$t$<\tau$\textsubscript{2}, the collapsed chains have not relaxed back into their original stretched configuration, which means the released \ce{K+} ions are still near the WE (region I), which adds more \ce{K+} ions to the total amount of mobilizable \ce{K+} ions: a PPF trend is observed instead. Only when $\Delta$t$>\tau$\textsubscript{2}, the system has had sufficient time to re-establish the brush configuration and ion distribution in regions I-II respectively, which yields neither facilitation nor depression: \textit{PPR(\%)} = 100\%.

From the paired-pulse experiments, we can derive that our polyelectrolyte brushes play a twofold role in the demonstration of memory-like behavior. The presence of a charged polymer brush influences the double layer formation and adds an additional capacitive contribution, which extends the retention time of ions near the WE. In our paired-pulse experiments, we see this ion retention as a paired-pulse \textit{depression} up to $\Delta$t-values of approx. 600 ms. The ion retention near the WE is much longer with PSPMA brushes present, as evident by the retention time measured on non-coated gold (50 ms). The charged nature of our polymer brushes also enables tunability of the synaptic behavior with the polarity of the trigger and with the salt concentration. Applying a potential with a polarity similar to the polyelectrolyte charge leads to minimal brush effects, while applying a potential with a polarity opposite to the polyelectrolyte charge may induce collapsing of PSPMA chains onto the gold surface and hereby releasing of condensed ions. Salt concentration also imparts the polymer coatings with different memory-effects, such as the suppression of current at short inter-pulse times (PPR $<$ 100\%, $\Delta$t $<$ 50 ms) and a facilitation of current in selected conditions (PPR $>$ 100\%, 0.01 M KCl, +0.2 V pulses, $\Delta$t = 50--600 ms). We attribute the occurrence of different memory effects to a difference in ion presence along the brush layer. Having shown that polyelectrolyte brushes can exhibit different memory-like behavior depending on the polarity and salt environment, we now focus on the accumulative charging within polymer brushes, which forms the basis for \textit{learning} and \textit{forgetting} behavior. 

\vspace{2cm}

\vfill

\clearpage

\begin{figure}[ht!]
    \centering
    \includegraphics[width=\linewidth]{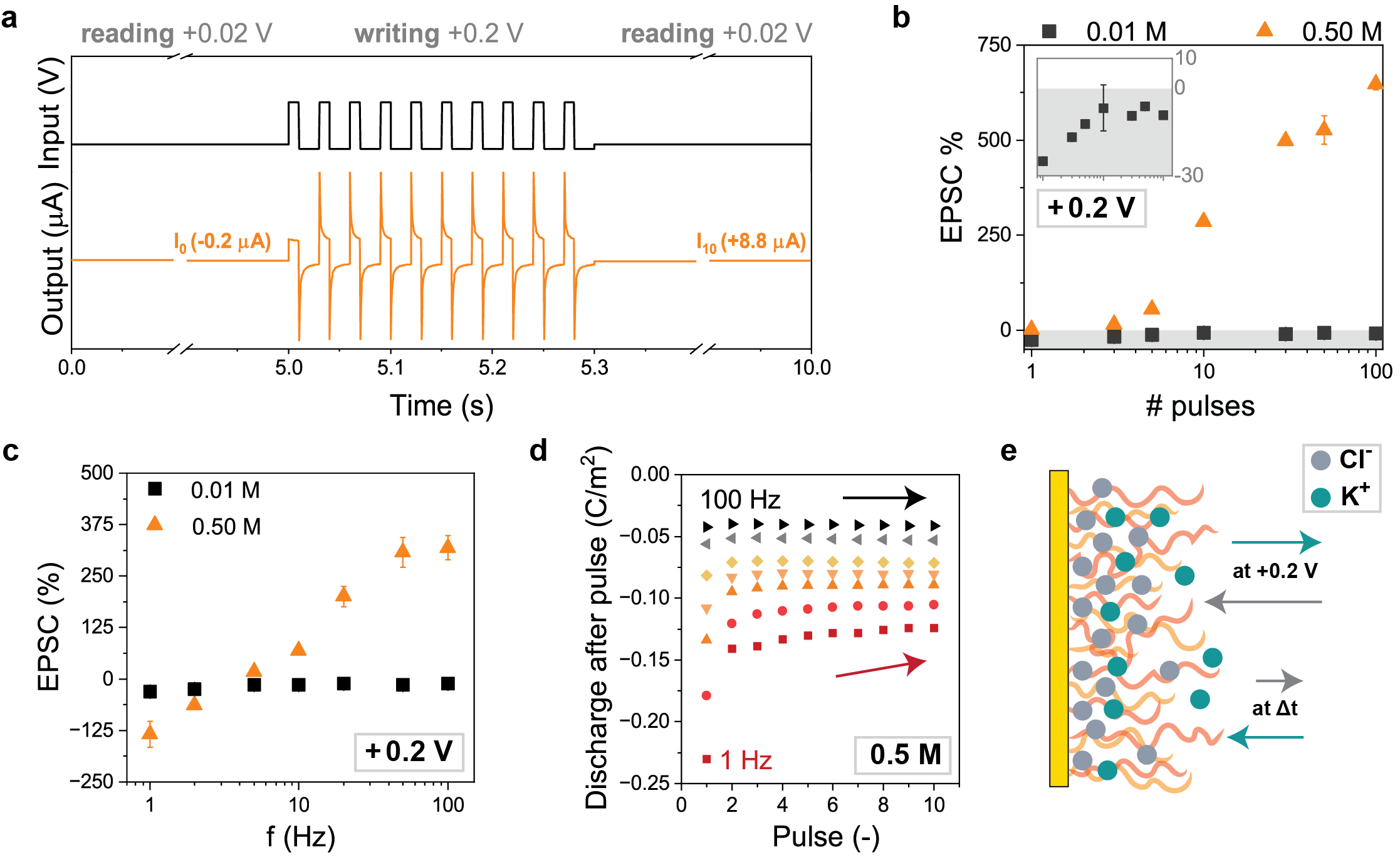}
    \caption{\textbf{Accumulative learning behavior.} a) Time trace of a SRDP experiment with \textit{f} = 50 Hz. b) SNDP tests performed with \textit{U} = +0.2 V in 0.01 M and 0.5 M KCl. Pulse numbers were varied between 1-100 pulses, while \textit{f} = 100 Hz. c) SRDP tests performed with \textit{U} = +0.2 V in 0.01 M and 0.5 M. Pulse frequencies were varied between 1-100 Hz, with \textit{\#pulses} = 10. d) Discharging of salted PSPMA brushes (0.5 M) during the pulse train. Color coded from 1 Hz (red) to 100 Hz (black). e) Mechanistic picture of the trend in (c), specifically the salted regime.}
    \label{fig:SRDP}
\end{figure}

Next, the short term plasticity of an artificial synapse was assessed by spike-number dependent plasticity (SNDP) and spike-rate dependent plasticity (SRDP) experiments. The \textit{learning} and \textit{forgetting} of biological synapses depends on the number and frequency (e.g. rate) of the incoming presynaptic signals.\cite{Cai.Networks.2014} The learning and forgetting of artificial synapses can be tested by subjecting the system to a pulse train of varying pulse numbers (here: \textit{\# pulses} = 1-100) and pulse frequencies (here: \textit{f} = 1-100 Hz). We express the change in current as the excitatory postsynaptic current (\textit{EPSC}), which is the relative current after the pulse train (\textit{I}\textsubscript{e}) compared to the current prior to the pulse train (\textit{I}\textsubscript{i}): \textit{EPSC\%} = $(I_e-I_i)/I_i \times 100\%$.  By reading the current prior to and after a pulse train, one can compare the number- and frequency-dependent \textit{learning} ($>$ 0\%) and \textit{forgetting} ($<$ 0\%) behavior of brushes in different conditions. An example of a pulse train experiment is shown in \textbf{Figure \ref{fig:SRDP}a}. In these experiments, we have used the same salt concentrations as in \textbf{Figure \ref{fig:PP}}, with pulses of 10 ms and \textit{U} = $\pm$ 0.2 V. The read voltage comprises 10\% of the pulse potential, so +20 mV for the +0.2 V pulses and -20 mV for the -0.2 V pulses. 
In \textbf{Figure \ref{fig:SRDP}b-c}, we present the SNDP (b) and SRDP (c) results with \textit{U} = +0.2 V, sorted by osmotic and salted conditions. 

The SNDP experiments in \textbf{Figure \ref{fig:SRDP}b} show a clear difference in the \textit{pulse number}-dependent learning behavior of PSPMA brushes between the osmotic and salted regime. With increasing pulse number, both salt regimes show an increase in EPSC\% in the range of 1-10 pulses, after which the EPSC\% either levels off (0.01 M) or continues increasing (0.5 M). Herein, the increase in the current signal is also significantly higher for salted brushes, which implies that a salted PSPMA brush is capable of 1) mobilizing more ions, and 2) charging its polymer matrix with more ions \textit{per pulse}, as compared to an osmotic PSPMA brush. The effect of charging will be further addressed in the section below. We attribute the difference in spike-number dependent plasticity to the screening of PSPMA charge centers in different salt concentrations, which cause a shift in ion-dynamics: in the osmotic regime, ion migration is regulated by the \textit{brush} (i.e. negative sulfonate moieties of SPMA), while in the salted regime, ions of the electrolyte are marginally affected by the charged moieties of the brush. Instead, the \textit{salted brush} forms a dense matrix in which the charges can be stored, and may allow for an overscreening effect,\cite{Sui.Macromol.2025} leading to EPSC\% values of 270--680 \% for 10-100 pulses. In a concurrent manner, we believe the stagnation of the EPSC-value in the osmotic regime stems from the fact that few mobile ions - primarily \ce{K+} counter-ions are present in the PSPMA brush, which hinders concentration of charges in the polymer matrix\cite{Qiu.JPhysChemC.2022} and thus building of EPSC\% each pulse. 

The SRDP experiments in \textbf{Figure \ref{fig:SRDP}c} show that salt concentration also significantly impacts the brush-related, \textit{frequency}-dependent learning behavior of our PSPMA brushes. Here, increasing the frequency of pulses increases the EPSC\% drastically for salted PSPMA brushes, and only slightly for their osmotic counterparts. We find that the variation in EPSC\% values in each salt regime is similar for both our SNDP (b) and SRDP (c) experiments. Based on the EPSC-values obtained during the SRDP experiments, we can categorize the frequency-dependent learning of our brushes into unipolar memory (0.01 M, + 0.2V) and bipolar memory (0.5 M, +0.2 V); a mechanistic explanation will be provided in the subsequent sections.

Frequency-dependent unipolar memory behavior is observed in the osmotic brush regime: the trend in EPSC\% shows frequency-dependency but is limited to \textit{forgetting} behavior ($<$ 0\%) throughout the entire frequency range. The negative EPSC\% values can the related to the discharging of the brush during the pulse trains (\textit{Supporting Information}, Figure \ref{fig:ChargeAcc}), which suggests that \ce{K+} is depleted from the PSPMA brush. Slow depletion of \ce{K+} to the outer regions of the PSPMA brush is osmotically driven by the presence of a double layer at the brush-electrolyte interface (Figure \ref{fig:DL}), which becomes significant when the salt concentration is low. The effect of such time-dependent, diffusional processes become more apparent at longer time frames -- e.g. large numbers of\textit{\#pulses} and low values of \textit{f} -- which explains the stagnating trend in \textbf{Figure \ref{fig:SRDP}b} and the negative EPSC\% values from \textbf{Figure \ref{fig:SRDP}c}.

Frequency-dependent bipolar memory is observed in the salted brush regime: the trend in EPSC\% transits from \textit{forgetting} ($<$0\%) to \textit{learning} ($>$0\%) with increasing pulse frequency. This synaptic plasticity behavior can be explained by a simultaneous effect of charging and brush relaxation. Charging becomes relevant when polyelectrolyte chains are largely ionized, and both mobile counter-ions (\ce{K+}) and co-ions (\ce{Cl-}) are present;\cite{Qiu.JPhysChemC.2022} an overscreening effect can occur at 0.5 M KCl (and not 0.01 M KCl).\cite{Sui.Macromol.2025} This is indeed visible in \textbf{Figure \ref{fig:SRDP}d} and Figures \ref{fig:Charge}--\ref{fig:ChargeAcc}, where we present the frequency-dependent discharging (d) and charging of the PSPMA brushes at 0.5 M KCl. At lower frequencies, when ion re-migration can occur between pulses  (f $<$ 1/$\tau_1$), accumulative charging is prevented and \ce{K+} is allowed to diffuse to the brush interface (Figure \ref{fig:DL}). The combination of these effects results in negative (for \textit{f} $<$ 10 Hz) and near-zero EPSC\% values (\textit{f $\sim$} 10 Hz). At higher pulse frequencies, when the pulse period is faster than the ion re-migration (f $>$ 1/$\tau_1$), we see that the time between pulses is not sufficient to reverse the charge separation from previous pulses. This is evident from the trend in discharging during pulses; little to no discharge is visible at \textit{f} = 100 Hz. As a result, each pulse in the pulse train contributes to an increasing extent of charge build-up, with the ions in the PSPMA brush building additional charge with each pulse. The overscreening effect -- which is illustrated in \textbf{Figure \ref{fig:SRDP}e} -- contributes to the increasing extent of charge separation and thus charge build-up. This charge separation eventually leads to an an increase in read current after the pulse train (\textit{I}\textsubscript{e}) and thus positive EPSC\% values. 

To compare these results with a different pulse polarity and with a reference substrate, we have included an overview in the \textit{Supporting Information} of SRDP experiments on a non-coated gold substrate (Figure \ref{fig:GoldSRDP}), and SRDP experiments using PSPMA brushes and \textit{U} = -0.2 V (S3.7). The non-coated surfaces display no learning or forgetting behavior in the frequency range of 1-100 Hz (EPSC $\sim$ 0\%), which suggests no accumulative effects can be attributed to the interactions of ions with the gold surface. At \textit{U}= - 0.2 V, little frequency-dependent behavior was observed in our PSPMA brushes, suggesting the occurrence of accumulative learning behavior in polyelectrolyte brushes is polarity-dependent.

\vspace{2cm}

\newpage
From the variations in spike number and spike frequency, we have derived that our polyelectrolyte brushes selectively display accumulative learning behavior. By comparing SRDP experiments on coated and non-coated surface, we determined that accumulative behavior only occurs in presence of a brush. The role of the brush is thus to facilitate accumulative processes, such as building of charges and charge separation in the polymer matrix. Similar to the paired-pulse experiments, we observed that our PSPMA brushes also enable tunability of this accumulative learning behavior with the polarity of the presynaptic signals and the salt concentration. For accumulative processes, applying a potential with polarity opposite to the polyelectrolyte charge is required. Salt concentration plays a role in the charging and discharging of the polymer coatings; salted brushes facilitate migration of both \ce{K+} counter-ions and \ce{Cl-} co-ions, which enables enhancement of the EPSC signal by an overscreening effect. Using 0.5 M KCl concentrations and positive polarity, we were able to vary the EPSC\% signal from +270\% to 680\% with increasing pulse number (SNDP at 100 Hz frequency), and from -125\% to +375\% with increasing frequencies (SRDP with 10 pulses). Having shown that polyelectrolyte brushes can exhibit different synaptic behavior depending on the polarity and salt environment allows us to understand how brush-regulated systems enable fine-tuning of iontronic memory devices. 

\newpage
\section{Conclusions}
In summary, we have shown that PSPMA brushes are capable of synaptic behavior in the most simple of electrochemical cell designs. Through experimental work in combination with supporting data of molecular dynamics (MD) simulations and mean-field approximations, we have unraveled the role of these charged brushes in the synapse-like response of these surfaces. In the first place, we show that there is a delayed retention of ions in the polymer brush compared to non-coated gold surfaces. This delayed retention is an effect of the brush itself, which adds a capacitive contribution to the electronic interface. As a result, the brush-functionalized surfaces exhibit paired-pulse depression in the sub-second range ($<$ 600 ms) and accumulative learning and forgetting depending on the presynaptic signalling, e.g. number of incoming signals and/or frequency of incoming signals. A secondary, yet very important, role of the polymer brush is to tune the interaction of ions with the polymeric layer, which can be done by changing the polarity of the presynaptic signal or by changing the salt concentration. Applying potentials opposite to the fixed charge on the brush enables number- and frequency-dependent learning behavior, as we have observed with SNDP and SRDP experiments. In addition, we see evidence for PSPMA chains collapsing at positive potentials, we contribute to a paired-pulse facilitation in low-salt conditions (0.01 M). Varying the salt concentration governs which and how many ions migration in the polymeric layer. Salted brushes in particular have the ability to build-up charges and, as a consequence, switch between learning and forgetting behavior. It is important to note that no such extreme dependencies on the polarity and ionic strength are found for non-coated surfaces, which strengthens the role polyelectrolyte brushes as a stimuli-responsive material in synaptic devices. The obtained trend in non-linear potential-current response, paired-pulse experiments and accumulative learning, helps advancing the design of polymer brush-based neuromorphic devices. 

\newpage
\section{Experimental}

\subsection{Polyelectrolyte brush fabrication}
The synthesis of 4-(2-hydroxyethyl)benzenediazonium and the attachment to gold was performed according to the procedure of Postma \textit{et al}.\cite{Postma.AdvMaterInt.2026} The synthesis and characterization of the PSPMA brushes proceeded according to earlier work,\cite{Postma.Macromol.2026,Smook.Macromol.2026} which is included in the \textit{Supporting Information}. Verification of the brushes' chemical composition, thickness and grafting density was done using (in-situ) ellipsometry, contact angle goniometry, polarization-modulated infrared reflection absorption spectroscopy (PM-IRRAS) and by grafting density estimations.

\subsection{Electrochemical characterization}
We compared SPMA-modified gold surfaces with electrochemical impedance spectroscopy (EIS) to indicate changes in the capacitive behavior of the WE when exposed to 0.01 M and 0.5 M KCl electrolyte. Chronoamperometry measurements were performed at $\pm$ 0.2 V to compare transient processes at the indicated voltages. Performed synaptic tests included paired-pulse experiments, spike-number dependent plasticity (SNDP) and spike-rate dependent plasticity (SRDP). Here, we measured our brushes in KCl electrolyte concentrations of 0.01 M and 0.5 M to create osmotically swollen and salted brushes, respectively. The chosen polarities were $\pm$ 0.2 V. Extensively methodologies are provided in the \textit{Supporting Information}.

\newpage
\begin{acknowledgement}

We thank Ivana Lin for initial discussions. We would like to acknowledge the financial support from NWO (KICH1.ST01.20.012), Andritz, Avebe, Corbion, Cosun, DSM-Firmenich, Meam, The Protein Brewery, and VNP. This publication is also part of Targets 1 and 5 of the TTW Perspectief research programme ReCoVR: Recovery and Circularity of Valuable Resources which is (partly) financed by the Dutch Research Council (NWO). This work used the Dutch national e-infrastructure with the support of the SURF Cooperative using grant no. EINF-9565.

\end{acknowledgement}

\begin{suppinfo}

The supporting information includes a step-by-step procedure of the PSPMA brush synthesis (S1), an overview of the characterization methods (S2) and supporting results (S3). The supporting results consist of a confirmation of the PSPMA brush synthesis, various electrochemical characterizations, reference datasets of paired-pulse and SRDP experiments on gold and coarse-grained molecular dynamics (GC-MD) simulations showing the charge distribution across the polymer brush.

\end{suppinfo}

\newpage
\bibliography{References}

\end{document}